%% file: main.tex
\documentclass[conference]{IEEEtran}

\usepackage{paco-notations}
\usepackage{supertech-llncs-light}
\usepackage{cleveref}
\usepackage{amsmath,amssymb,amsfonts}
\usepackage{algorithmic}
\usepackage{graphicx}
\usepackage{textcomp}
\usepackage{xcolor}
\def\BibTeX{{\rm B\kern-.05em{\sc i\kern-.025em b}\kern-.08em
    T\kern-.1667em\lower.7ex\hbox{E}\kern-.125emX}}
    
\begin{document}
\title{A Reexamination of the Communication Bandwidth Cost Analysis of A Parallel Recursive Algorithm for Solving Triangular Systems of Linear Equations}

\punt{% begin punt
\author{Yuan Tang}\thanks{Corresponding Author}
\authornote{%
Also affiliated with Shanghai Key Lab. of Intelligent Information Processing.}
\affiliation{%
    \institution{School of Computer Science, School of Software, Fudan University}
    \city{Shanghai}
    \country{P. R. China}
}
\email{yuantang@fudan.edu.cn}
}% end punt

% \punt{% begin punt
\author{\IEEEauthorblockN{Yuan Tang}
\IEEEauthorblockA{\textit{School of Computer Science, School of Software} \\
\textit{Fudan University}\\
Shanghai, P.R.China \\
yuantang@fudan.edu.cn}
}
% }% end punt

\maketitle
\input{abstract}
% \input{appendix}
%\clearpage
%\newpage
% \input{intro}
\input{trsm}
% \input{concl}

% \clearpage
% \newpage
% \bibliographystyle{ACM-Reference-Format}
% \bibliographystyle{abbrv}
\bibliographystyle{IEEEtran}
\bibliography{papers}

% \clearpage
% \appendix
% \input{symbol-table}
% \input{p-balanced-model}
% \input{lcs}
% \input{sort}
% \input{expr}
% \input{suppl}
	
%\printbibliography
\end{document}

%% file: abstract.tex
\begin{abstract}

% State the problem
% Why it's an interesting problem
% What your solution achieves
% What follows from your solution: Impact of your solution
This paper presents a reexamination of the research paper titled ``Communication-Avoiding Parallel Algorithms for \proc{TRSM}'' by Wicky et al. We focus on the communication bandwidth cost analysis presented in the original work and identify potential issues that require clarification or revision. The problem at hand is the need to address inconsistencies and miscalculations found in the analysis, particularly in the categorization of costs into three scenarios based on the relationship between matrix dimensions and processor count. Our findings contribute to the ongoing discourse in the field and pave the way for further improvements in this area of research.
\end{abstract}
% \punt{% begin punt
% \keywords{
\begin{IEEEkeywords}
Communication-Avoiding Parallel Algorithm,
communication bandwidth complexity,
Triangular Systems of Linear Equations,
TRSM
% }
\end{IEEEkeywords}
% \end{IEEEkeywords}

% }% end punt

%% file: trsm.tex
\secput{intro}{Introduction}
Wicky et al. \cite{WickySoHo17} proposed an innovative inversion-based parallel algorithm for triangular solve with multiple right-hand sides (\proc{TRSM} in the LAPACK terminology) to minimize communication costs. Their algorithm builds upon a recursive \proc{TRSM} algorithm. However, upon closer examination, we have identified several potential issues with the theoretical analysis that require further investigation. Specifically, we focus on the categorization of communication costs based on the relationship between matrix dimensions and the number of processors.

\section{Background}
We begin by providing pertinent background information on the \proc{TRSM} algorithm and the cost model used by Wicky et al. \cite{WickySoHo17}. The objective is to solve the equation $LX = B$, where $L \in \mathbb{R}^{n\times n}$ is a lower triangular matrix, and $B \in \mathbb{R}^{n\times k}$ contains $k$ right-hand sides. Their recursive algorithm (prior to their final inversion-based approach) recursively partitions $L$ and $B$ until a base case size $n_0$ is reached or the processor count $p = 1$. The overall cost is analyzed in terms of the floating-point operation (flop) count $F$, the bandwidth $W$ (number of words of data sent and received), and the latency $S$ (number of messages sent and received) along a critical path \cite{SolomonikCaKn14}, denoted by $T = \alpha \cdot S + \beta \cdot W + \gamma \cdot F$, where $T$ represents the total execution time, and $\alpha, \beta, \gamma$ denote the unit cost of latency, bandwidth, and floating-point operation, respectively.

\section{Limitations of Original Analysis}

    \begin{figure}[htbp]
		\centering
			\fbox{\includegraphics[width=\linewidth]{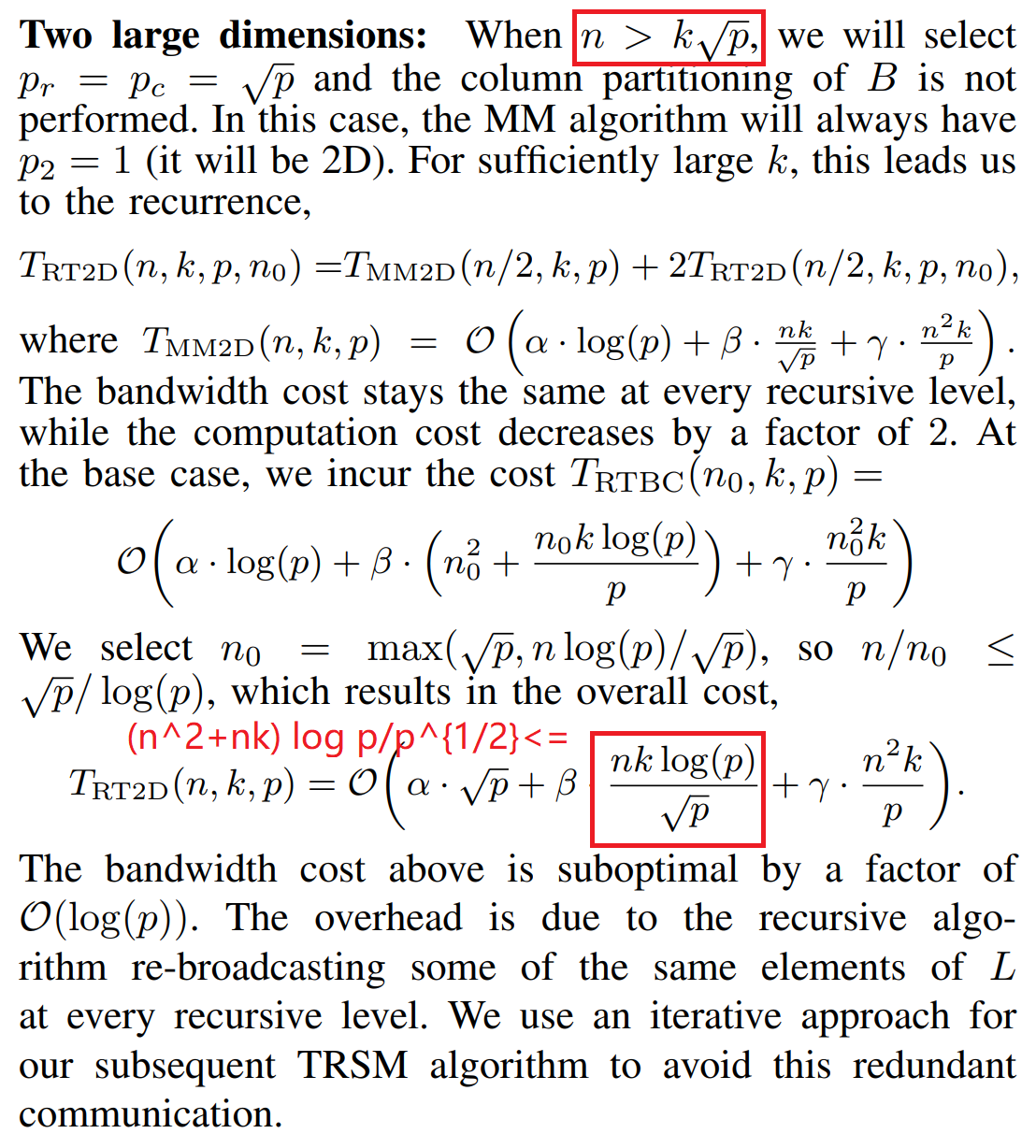}}
			 \caption{The problematic condition of $n > k\sqrt{p}$ for ``two large dimensions'' exhibits gaps; The bandwidth bound associated with $\beta$ should be $\frac{(n^2 + nk) \log p}{\sqrt{p}}$.}
			 \label{fig:two-large-dims}
	\end{figure}
 
     \begin{figure}[htbp]
		\centering
			\fbox{\includegraphics[width=\linewidth]{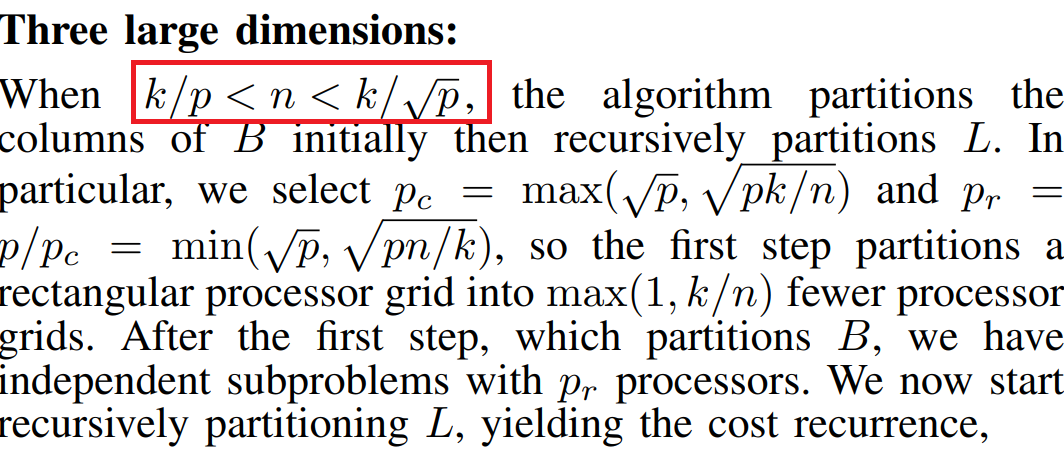}}
			 \caption{The problematic condition of $k/p < n < k/\sqrt{p}$ for ``three large dimensions'' exhibits gaps.}
			 \label{fig:three-large-dims-cond}
    \end{figure}

    \begin{figure}[htbp]
        \centering
			\fbox{\includegraphics[width=\linewidth]{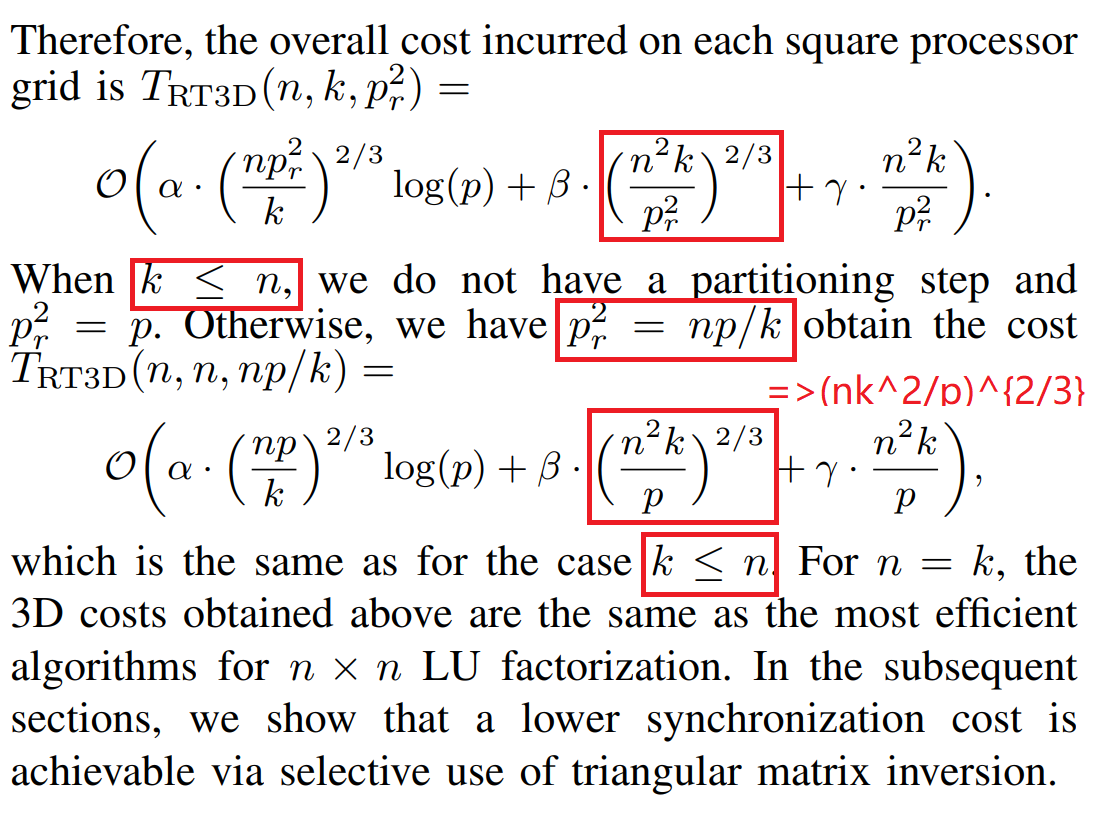}}
			 \caption{The incorrect bandwidth bound of ``three large dimensions'' : with the proposed $p_r^2 = np/k$, the bandwidth bound associated with $\beta$ should be $(\frac{nk^2}{p})^{2/3}$}
			 \label{fig:three-large-dims-bound}       
    \end{figure}

\punt{% begin punt
The original analysis \cite{WickySoHo17} classifies the costs of the recursive algorithm into three cases, which depend on the relationship between $n$, $k$, and the number of processors $p$. These cases are as follows: (i) one large dimension ($n < k/p$); (ii) two large dimensions ($n > k\sqrt{p}$); and (iii) three large dimensions ($k/p < n < k/\sqrt{p}$). This categorization follows that of the rectangular matrix multiplication algorithm described in the CARMA paper \cite{DemmelElFo13}. However, we will demonstrate that this categorization may have certain issues.
}% end punt

\begin{enumerate}
\item In the original Section IV A, the authors categorized the cost analysis of their recursive \proc{TRSM} algorithm into three categories, drawing an analogy to the CARMA paper \cite{DemmelElFo13} for rectangular matrix multiplication. However, this categorization may be inappropriate because the \proc{TRSM} problem for rectangular matrices has only two free dimensions to scale (namely, $n$ and $k$), unlike the three independent dimensions ($m, n, k$) of rectangular matrix multiplication.

\item In the case of two large dimensions (refer to \figref{two-large-dims}), the condition $n > k\sqrt{p}$ exhibits gaps in coverage within the 2D scope. Specifically, when $1/\sqrt{p} < k/n < \sqrt{p}$, the costs are undefined, regardless of the three cases: one large dimension ($n < k/p$), two large dimensions ($n >k\sqrt{p}$), and three large dimensions ($k/p < n < k/\sqrt{p}$).

Furthermore, we have received valuable insight through personal communication with Prof. Solomonik \footnote{One of the coauthors of the original paper}, suggesting that the condition for two large dimensions may need to be revised to $n < k\sqrt{p}$. However, even with this modification, two issues arise:
    \begin{enumerate}
        \item A gap emerges when $k/n < 1/\sqrt{p}$.
        
        \item When $n =\Omega(k)$, the bandwidth cost increases to $O((nk + n^2)\log(p)/\sqrt{p})$, potentially surpassing the stated claim of $O(nk\log(p)/\sqrt{p})$ in the asymptotic sense.
    \end{enumerate}

\item In the case of three large dimensions (refer to \figref{three-large-dims-cond}), the condition $k/p < n < k/\sqrt{p}$, which is equivalent to $\sqrt{p} < k/n < p$, contradicts the possibility of $k \leq n$ discussed in the last two paragraphs of this case (refer to \figref{three-large-dims-bound}). This contradiction arises because $k > \sqrt{p}n$ in this scenario, and $\sqrt{p} \geq 1$.

By incorporating the suggested $p_r^2 = np/k$ (refer to \figref{three-large-dims-bound}), the bandwidth cost becomes $O(\beta(nk^2/p)^{2/3})$ when $k > n\sqrt{p}$, once again exceeding the paper's claim of $O(\beta(n^2k/p)^{2/3})$, considering that $k > n\sqrt{p}$.
\end{enumerate}

In summary, the categorization of \proc{TRSM} into three cases leads to significant logical inconsistencies, gaps in coverage, and inaccurate cost formulas. A revision may be required.

\secput{concl}{Conclusion}
In this paper, we have reexamined the communication bandwidth cost analysis of the recursive \proc{TRSM} algorithm from \cite{WickySoHo17}. Our investigation has revealed limitations in the original formulation and categorization of costs based on matrix dimensions and processor count. Specifically, the three-case categorization, derived from the CARMA paper \cite{DemmelElFo13}, does not adequately consider the constraints of \proc{TRSM}, due to its fewer degrees of freedom. Our reanalysis has highlighted problematic gaps in coverage and logical inconsistencies. We hope that our findings will inspire further advancements in this field.

\punt{% begin punt
In conclusion, this work contributes to the refinement and strengthening of the underlying asymptotic complexity theory for an important class of numerical linear algebra algorithms. We emphasize the importance of rigorous reevaluation of published findings, particularly with regards to algorithmic cost models. Continued reexaminations will ensure the robust advancement of knowledge in the field.
}% end punt

\secput{ack}{Acknowledgement}
We would like to thank the authors of the original paper for their contributions to the field, and we appreciate the opportunity to discuss and critique their work. Our intention is not to diminish the value of their research, but rather to contribute to the ongoing scientific discourse and to the shared goal of improving the performance and efficiency of parallel numerical algorithms. 